# A possible role of sunrise/sunset azimuth in the planning of ancient Chinese towns

**Amelia Carolina Sparavigna**
**Department of Applied Science and Technology, Politecnico di Torino, Italy**

In the planning of some Chinese towns we can see an evident orientation with the cardinal north-south direction. However, other features reveal a possible orientation with the sunrise/sunset azimuth on solstices too, as in the case of Shangdu (Xanadu), the summer capital of Kublai Khan. Here we discuss some other examples of a possible solar orientation in the planning of ancient towns. We will analyze the plans of Xi'an, Khanbalik and Dali.

In a recent paper I have discussed a possible solar orientation of Shangdu, also known as Xanadu, the summer capital of Kublai Khan [1]. Xanadu has an evident north-south orientation; however, the remains of the walls seem to have some features, planned according to the direction of sunrise and sunset on the solstices, as we can easily see using Google satellite images and a software, provided by sollumis.com, which gives the solar azimuth during the year. Some solar orientations can also be found in the planning of Chinese Pyramids burial complexes, as discussed in the Reference 2.

The town of Xanadu was planned by the architect Liu Bingzhong (1216–1274), the Kublai's adviser. According to Reference 3, the city was designed implementing the traditional scheme for the city's architecture. The ancient Chinese urban planning included some symbolisms concerning cosmology, geomancy, astrology, and numerology [4], with the aim of maintaining a local harmony and balance. One of the most evident features of this urban planning is a grid having a north-south orientation [4]. As told in [3], this orientation was already present in a very old Neolithic settlement, found at Banpo, near Xi'an. This settlement is representing what was probably a typical settlement of North China, five or six thousand years ago [3].

Let us note that a north-south orientation is different from a solar orientation with sunrise/sunset azimuths. The north-south direction is the projection on the Earth surface of the "axis mundi", which is the cosmic axis connecting Heaven and Earth and passing through the celestial poles. This direction can be easily found using the shadow of a gnomon, as discussed by Vitruvius [8,9], or by the observation of the rising and setting of some stars [10].

Orienting temples, monuments or towns due North could mean an orientation with the axis about which all the world is rotating (to ancient people the world was the physical universe). In the case of a solar orientation, the plan of buildings or towns has a direction usually inclined with respect the cardinal direction, because the solar azimuth changes during the year. However, the two orientations can be combined: in the planning of Xanadu we can find the north-south orientation with Heavens, and also an orientation with the solar azimuth on solstices. As we can see from Figure 1 for instance, from the center of the imperial palace, looking at the south gate, on the winter solstice, the sun was rising and setting in correspondence of two bastions of the walls. Then, the orientation according to the solar azimuth is obtained by means of some elements of the walls, which become like the boundary of a local symbolic horizon.

In 1276, acting on the earlier advice of Liu Bingzhong, Kublai Khan decided to reform the calendar and therefore several observatories were built throughout China in order to perform the astronomical measurements suitable to the calendar revision [1]. We could tell that the Khan and his architect had a common deep interest in astronomy.

The orientation of buildings to cardinal directions is considered one of the general characteristics of the Chinese imperial urbanism [3], such as the regular subdivision of the urban plan. This fact was remarked by Francis John Haverfield [5]: in his book on the planning of ancient towns, he proposed that the Chinese towns were laid out in a fashion connected with a very old agrarian system. In fact, a link between town-planning and agrarian system can also be found for the ancient Roman towns, whose planning had based on the "centuriation", a method of land measurement and subdivision. The surveyor first determines a central viewpoint and then looking towards the rising sun, orients the land and subdivides it. With this method, the town results with a solar orientation [6,7]. And therefore, several Roman colonies have the main street, the decumanus, oriented with the sunrise azimuth on the day of their foundation. Then, a pure solar orientation is given according to the "orientem", that is the rising sun, whereas an orientation with the cardinal north-south direction is according to the "cardo", the pivot on which the world turns, that is, the pole of the sky.

F.J. Haversfield is dating back the Chinese town-planning to two or three thousand years ago. But, one of its features, the north-south orientation, is quite older as demonstrated by the Banpu settlement. In Ref.5 or 3, it is not discussed an orientation of ancient Chinese town-planning according to the rising and setting of the sun. Here we show some examples of solar orientation in the urban plan of Xi'an, Khanbalik and Dali. Xi'an and Khanbalik are cases where, as we have already found in Xanadu, there is a possible solar orientation combined with the cardinal orientation. However, before discussing them, let me shortly report some part of the Appendix written by Haverfield in his book [5].

**Haverfield's discussion on the Chinese town-planning**

Haverfield tells that *Many towns in China and also in Japan show more or less definite traces of 'chess-board' planning which recall the customs of the Macedonian and Roman worlds. The outlines of such towns are sometimes rectangular, though sometimes wholly irregular as if the sport of local conditions; their streets, or at least their main streets run generally straight and at right angles to one another and end at symmetrically placed gateways. This is no modern device. Probably it goes back two thousand, or even three thousand, years. The illustration which I* (F.J. Haverfield) *give here, in fig. 36* (the map is reproduced in Fig.2, where it is compared with satellite Google Maps)*, showing one of the Chinese military colonies planted in Turkestan in the eleventh or twelfth century, is selected not as the oldest, but as the best example which I can find of more or less ancient Chinese planning. There seems no doubt that the system itself is very much more ancient than this instance. Even in Japan, which probably copied China in this respect, towns were laid out in chess-board fashion long before the twelfth century; such are The former capitals, Kioto and Nara, the latter of which is said to have been founded as early as A.D. 708.*

*Probably the custom is connected with a very old agrarian system, sometimes known as the 'Tsing' system (from the shape of the Chinese character of that name), according to*

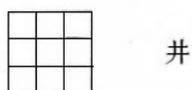

*which land was divided into square parcels and each parcel was subdivided into nine equal squares. The origin of this system has been ascribed to the twelfth and even to the eighteenth century B.C. ... It seems that this system may be closely connected with the system of laying out settlements and towns, which developed collaterally with it and produced Chinese town-*

*planning. In China, as at Rome, it would appear that the technical principles on which town and country were laid out were intimately akin. One item in the Chinese 'chess-board' plan is curiously parallel to a feature which occasionally occurs in Roman towns. In many Chinese cities, where the streets are straight and run at right angles to one another, the gates towards which they point are nevertheless not vis-à-vis, but the main thoroughfares between the gates make two right-angled turns at some point in their otherwise straight course. Thus travellers do not pass through the town in one continuous straight line, and, as in fig. 36, the east gate is not just opposite the west gate but a short distance to the right or left. ... For ordinary European readers this town-planning of ancient China raises the interesting, if unprofitable, question of the intercourse between ancient China and ancient Europe. The Chinese scholar Terrien de la Couperie once suggested that Chinese civilization was derived from Babylonia somewhere about 2600 B.C. Arguments concerning that period are perhaps dangerous, but it is hardly probable that the Babylonians used town-planning so long ago. Whatever connexion there may have been in ancient times between eastern and western town-planning must be more recent. Our knowledge of these more recent times, however, does not make a connexion seem probable. Not only are the physical obstacles plain enough — the long perilous seas that wash the coasts of southern Asia and the immense deserts and immense mountains of the central continent- but satisfactory proof of real intercourse is wanting. If, as one recent writer emphasizes, the spheres of Roman and Chinese influence, about A.D. 100, were divided only by the waters of the Caspian, if Chinese silk was brought to Rome, and two or three embassies set out from Rome to China or from China to Rome, nothing happened in consequence. Chinese coins are unknown in the Roman Empire, and Roman coins are exceedingly rare in China or anywhere east of Cape Comorin. The Roman geographers, who knew the coasts of Asia to Ceylon, knew little beyond it, and Ptolemy failed to see that the overland route to the far east and the sea-route led to the same country. The early Chinese had no doubt heard of the Roman Empire, just as the Romans had heard of China; such hearsay does not produce any great effect on the civilization of either side. The Chinese doubtless knew much more of Bactria; and Greek, or rather Graeco-Buddhist art, has left abundant traces in the desert cities of central Asia as far as the Chinese Wall. But a town-plan is too complex a thing to travel well. It is plainly more likely that east and west reached their similar results quite independently. ...*

The site shown is the Figure 2 is that of Khara-Khoto (Eji Nai City or Heishui City). According to [11], the town was founded in 1032 and became prosper trading center in the 11th century. There are remains of high and thick outer walls. Inside there was a walled fortress, first taken by Genghis Khan in 1226; the city continued to flourish under Mongol rulers. In the Figure 3 we can easily see that the town was larger than its citadel (the satellite image was processed as discussed in [12,13]). Kublai Khan ruling, the city was expanded, reaching a size three times larger the original one. The city was located on the crossroads connecting Karakorum (see Figure 4), Xanadu (Figure 1) and Kumul. Marco Polo visited a city called *Etzina* or *Edzina*, which has been identified with Khara-Khoto [11].

From Figure 2 and 3, we see that Khara-Khoto has an axis on the east-west direction. Karakorum has not the streets oriented with the east-west cardinal direction, but with an azimuth angle of approximately 115°. This angle corresponds to the azimuth of sunrise on the beginning of February or November. It is, more or less, the same orientation of Torino [6].

**Xi'an**
Xi'an has a rich and culturally significant history. As previously told, the 6,500 year old Banpo Neolithic village was discovered on the outskirts of the city [14,15]. Xi'an became a political center of China in the 11th century BCE, with the Zhou Dynasty. The capital of Zhou was established in the twin settlements of Fengjing and Haojing, located southwest of contemporary Xi'an. After the Warring States Period, China was unified under the Qin Dynasty (221-206 BCE) for the first time, with the capital located at Xianyang, northwest the modern Xi'an. The first emperor of China, Qin Shi Huang has his mausoleum near Xi'an.

In 202 BCE, the emperor Liu Bang of the Han Dynasty established his capital in Chang'an County, and this is traditionally considered the founding date of Chang'an, or Xi'an. The original Xi'an city walls creation started in 194 BCE and took 4 years to be completed. The walls measured 25.7 km in length and 12–16 m in thickness at the base. Chang'an was devastated at the end of the Tang Dynasty in 904 CE and the residents were forced to move to the new capital Luoyang. The town-planning of Chang'an during the Tan dynasty is given in Ref.16. It is possible to see the subdivision of the land, according to the rules discussed in [4,5]. During the Ming Dynasty, new walls were constructed in 1370 and we can see them intact today. The walls measure 11.9 km in length, 12 m and 15–18 m in thickness at the base; a moat was also built outside the perimeter.

Xi'an has possibly both cardinal and solar orientations. Using the satellite maps, we can easily see that the town has a main street oriented with the cardinal north-south axis (see Fig.5), and the walls have two gates on this axis. The walls have east and west gates, which are on a perpendicular street of course; however, if we use sollumis.com software, as we did for Xanadu, we see that these gates can be seen from the crossing of two main roads with the azimuths of sunrise and sunset on winter solstice. It seems then that, together with the cardinal orientation, the east and west gates were arranged according to the azimuth of the sun on the winter solstice. Again, the walls could represent the boundary of a local symbolic horizon.

Here, it is suitable to report what Ref.17 is telling on the grid-plan of Chinese towns, about the guidelines written in Kaogongji during the Spring and Autumn Period (770-476 BC): "a capital city should be square on plan. Three gates on each side of the perimeter lead into the nine main streets that crisscross the city and define its grid-pattern. And for its layout the city should have the Royal Court situated in the south, the Marketplace in the north, the Imperial Ancestral Temple in the east and the Altar to the Gods of Land and Grain in the west."

**Khanbalik**

Khanbalik or Dadu refers to a city which is now Beijing. The city was called Dadu (大都, pinyin), meaning "great capital" or "grand capital" of the Yuan Dynasty founded by Kublai Khan, It is known as Khanbalik (汗八里), meaning the "great residence of the Khan", and Marco Polo wrote of it as Cambaluc [18]. The architect and planner of Dadu was Liu Bingzhong, the architect of Xanadu. The construction of the walls of the city began in 1264, while the imperial palace was built from 1274 onwards. According to [18], the design of Dadu followed the rules of 9 vertical axes and 9 horizontal axes, "palaces in the front, markets in the rear", "left ancestral worship, right god worship" were taken into consideration. And this seems the subdivision of the ancient Roman towns, during the "centuriation" [6].

The perimeter of Khanbalik is shown in the Figure 6. This town has a north-south cardinal orientation. However, has it a solar orientation included? Using the sollumis.com diagrams for the winter and the summer solstice, we can find two interesting facts, depicted in Figs.7 and 8. Let us look at Figure 7: there we have a polar diagram for the winter solstice, having the origin at the crossing of two main roads in the northern part of the city. Using an origin on the Imperial Garden of Forbidden City, we can have a polar diagram for the summer solstice. The winter and summer sunrise/sunset azimuthal directions cross at two points of the perimeter, which determine a line halving the city. Moreover, in the Figure 8, we can see that using the same origins for the polar diagrams for the direction of the sun during the summer/winter solstice, the azimuth of sunrise and sunset is passing through the corners of the perimeter.

And this seems to be a solar orientation nested in the cardinal planning of the town, to represent a local horizon.

Inside Khanbalik , we have The Forbidden City. It was the Chinese imperial palace from the Ming Dynasty to the end of the Qing Dynasty, built from 1406 to 1420, and consisting of 980 buildings [19]. It is remarkable the bilateral symmetry of the plan of these buildings. This symmetry signifies balance [20]. And we have the symmetry of the sun direction too, as shown in the Figure 9. The polar diagrams have origin at the Hall of the Central Harmony in the Forbidden City, which is between the Hall of Preserving Harmony and that of the Supreme Harmony. The azimuths of sunrise and sunset are passing through the corners of the perimeter of the structure. Then, observing the sun from the Central Hall during the year, we have a balance of solar directions.

These examples proposed in Fig.7,8 and 9 show that the bilateral symmetry, used in the planning of a town had as a consequence a symmetry of solar directions, in the case the town had a cardinal orientation. It is then quite possible that architects included also, in their planning, an orientation with the sunrise/sunset azimuths on solstices.

However, the ancient Chinese towns are not only planned on the cardinal orientation. Let us see a good example of pure solar orientation, as we can find for the Roman towns [6,7].

**Dali and the sun**

Dali is a town in the Yunnan province of China. It was the ancient capital of both the Bai kingdom Nanzhao, which flourished in the area during the 8th and 9th centuries, and the Kingdom of Dali, which reigned from 937–1253. Founded in 937, in 1253 it was annexed in the Mongol Empire. The Kingdom of Dali was preceded by the Nanzhao Dynasty, which was overthrown in 902. Three dynasties followed in quick succession, until Duan Siping became king in 937, founding Dali. The 11th king of Nanzhao established Buddhism as the state religion: several of the kings of Dali gave up the throne and became monks [22,23]. In 1274 the Province of Yunnan was created by the Yuan Dynasty.

The old city was built during Ming Dynasty emperor Hongwu's reign (1368–1398) [22]. Figure 10 shows it, as we can see in Wikimapia, with its walls and gates. We have north and south gates, and east and "mountain" gates. Note that the town has not a cardinal orientation. Using sollumis.com, we have the result shown in the Figure 11: the direction of the streets is the same as that of the sunrise azimuth on 20 of May (or July), more or less. Therefore, a "pure" solar orientation of the town-planning is possible.

**Conclusions**

In these examples we have seen, that besides the cardinal orientation of the town-planning, we can also have some elements showing a solar orientation, that is, an orientation according to the azimuth of sunrise/sunset on solstices. However, other towns can have a pure solar orientation of their streets, as that we can observe for the ancient Roman towns. It seems then that an investigation of many Chinese towns, using satellite maps and software for solar analysis, could be interesting to obtain a more statistically significant result.

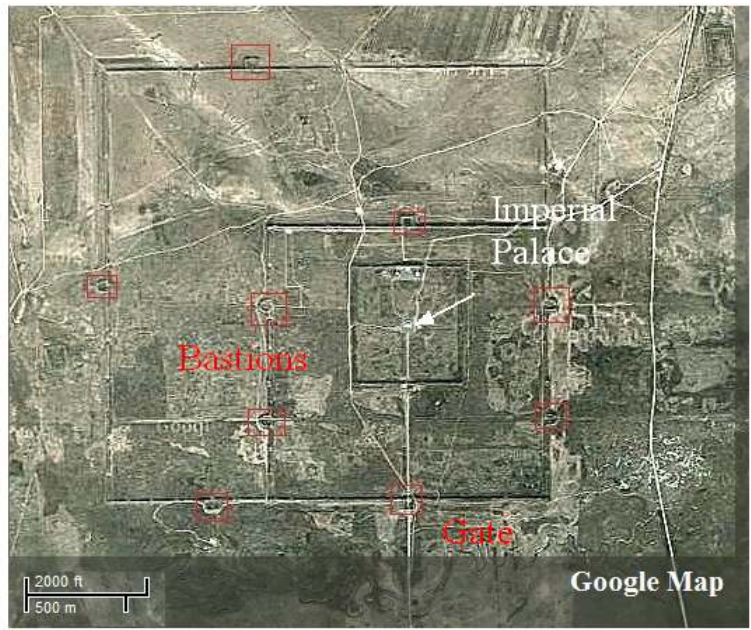

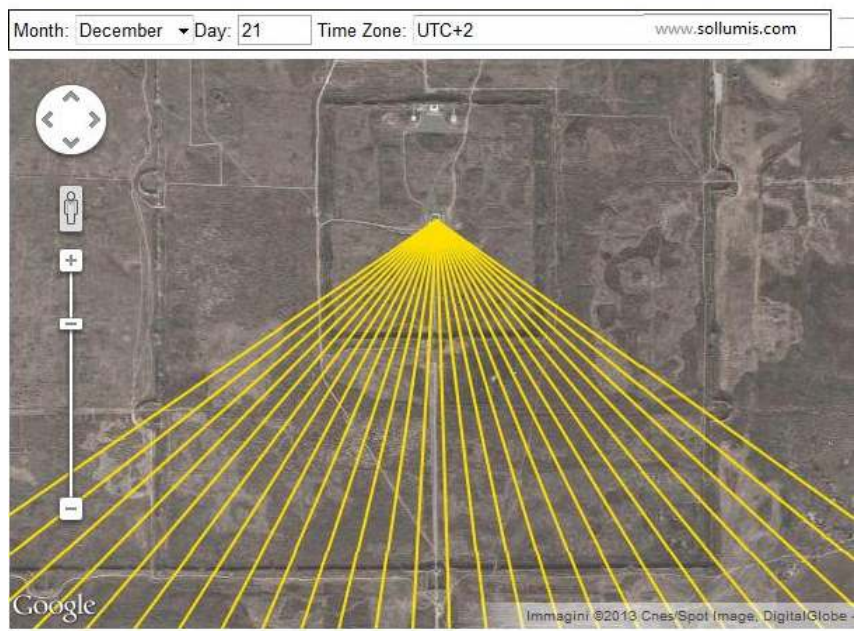

Figure 1: The ruins of Xanadu from the Google Maps. The walls have two gates oriented north-south and some circular protruding structures, probably round bastions, that in the image are marked in red. From the center of the imperial palace, looking at the south gate, on the winter solstice, the sun was rising and setting in correspondence of two bastions of the walls. This orientation according to the solar azimuth is rendering the walls like the limit of a local symbolic horizon (probably Kublai Khan and his architect shared a common interest in astronomy). The direction of the sun during the winter solstice is given by Sollumis.com (http://www.sollumis.com/). This site provides a polar diagram, overlaying a satellite map, showing the directions of the sun during any day of the year. The lines on the drawing show the direction and height (altitude) of the sun. Thicker and shorter lines mean the sun is higher in the sky. Longer and thinner lines mean the sun is closer to the horizon.

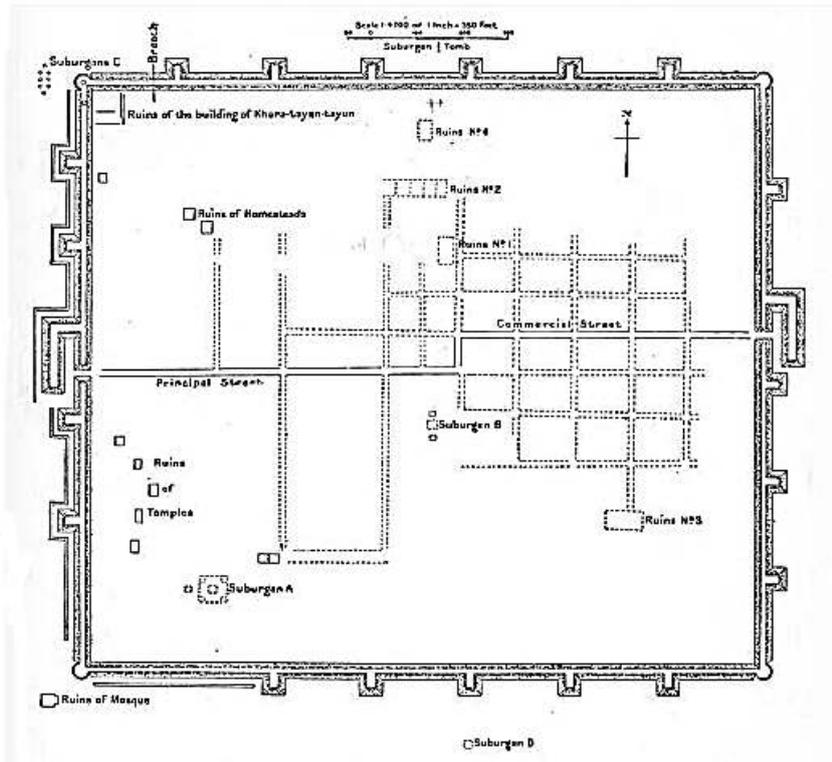

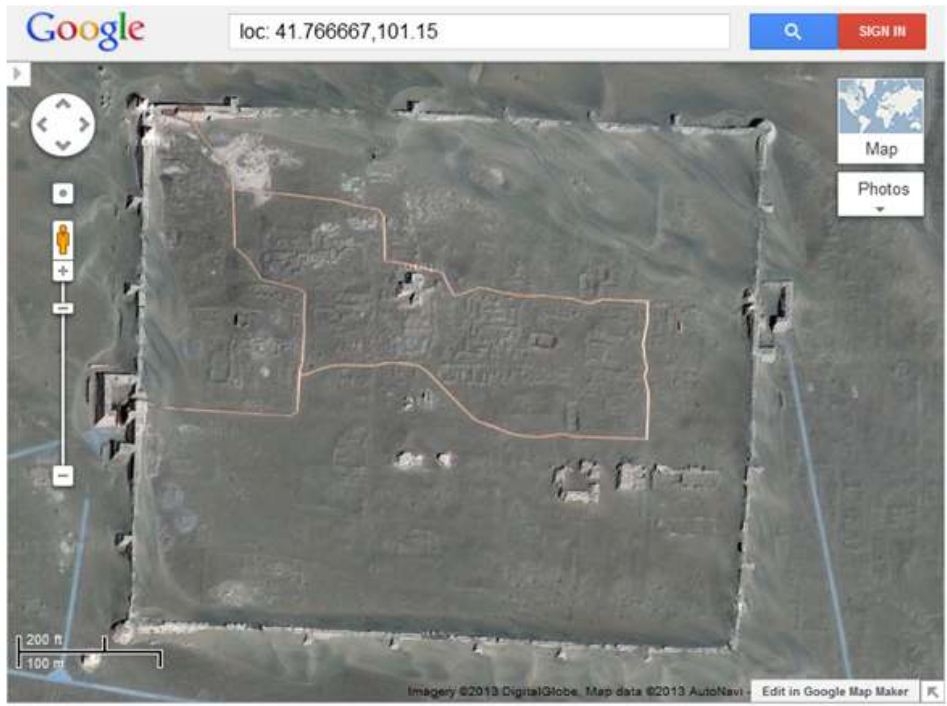

Figure 2: In the upper part it is shown the illustration given by F. Haverfield of Khara-Khoto. In the lower part of the image, the Google Map of the site.

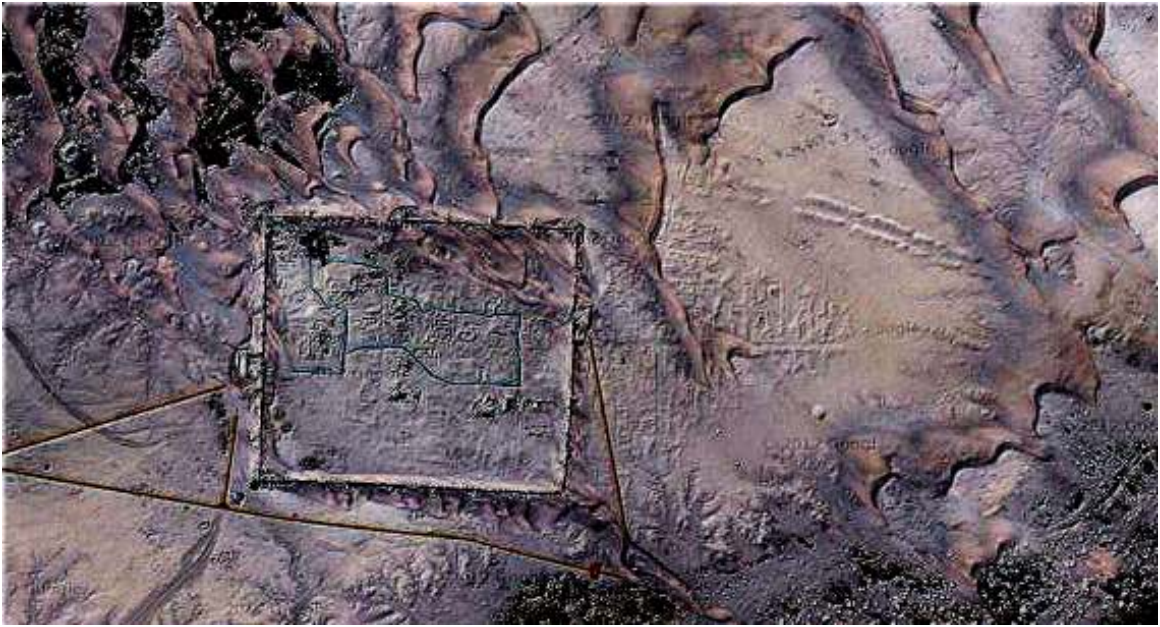

Figure 3: Khara-Khoto was larger than its fortified citadel, as we can see in the satellite maps. The map has been processed using AstroFracTool and GIMP [12].

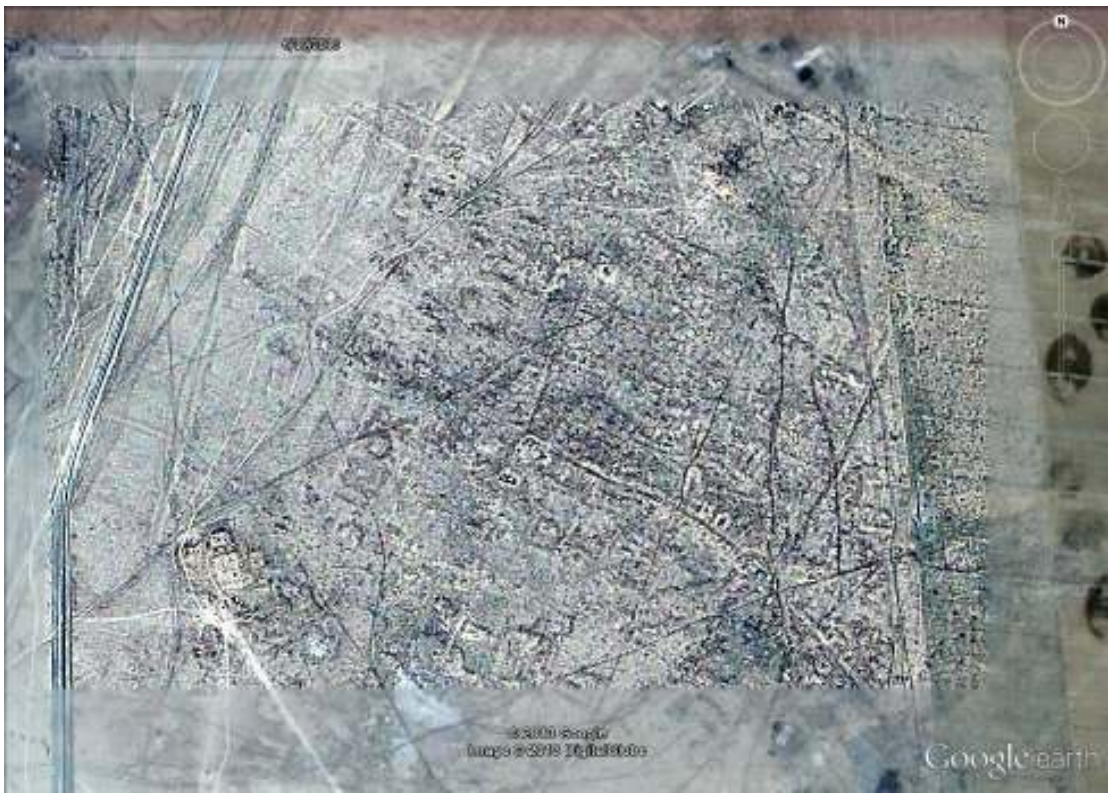

Figure 4: The site of Karakorum as given by Google Earth, after a processing [13]. The streets have not an orientation with the east-west cardinal direction, but they have an azimuth angle of approximately 115°, corresponding to the azimuth of sunrise on the beginning of February or November. It is more or less, the same orientation of Torino [6].

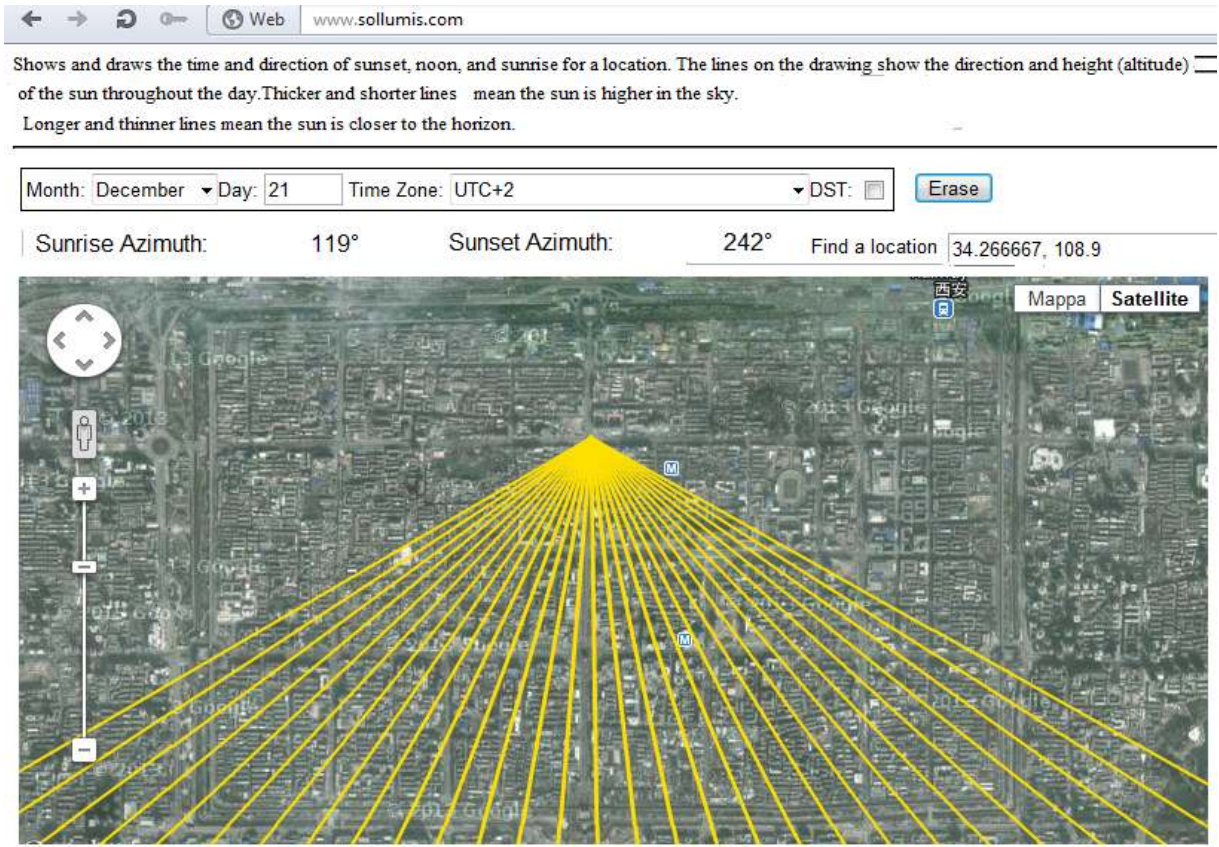

Figure 5: The east and west gates of Xi'an can be seen from the crossing of two main roads with the azimuth of sunrise and sunset on the winter solstice. We can easily see it by means of the polar diagram at the site sollumis.com on the Google Maps.

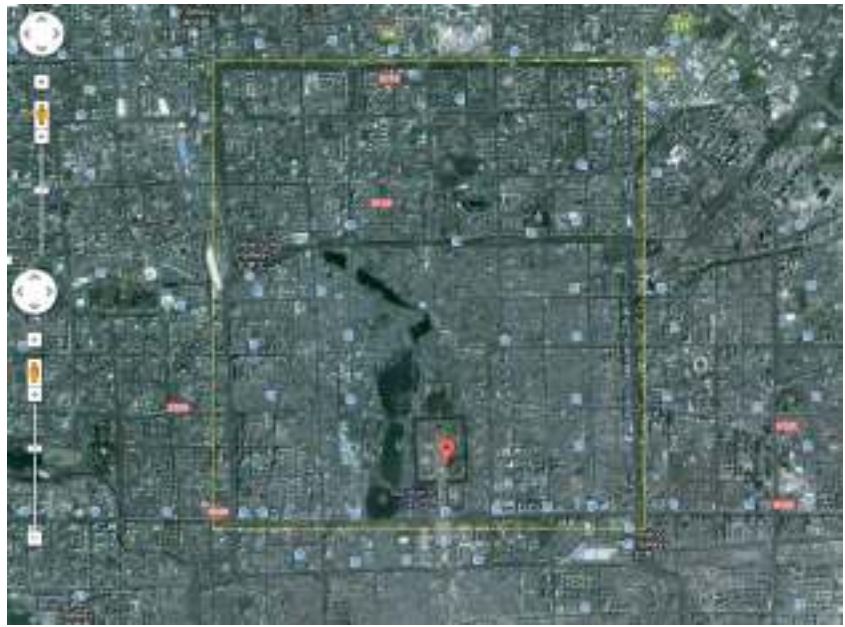

Figure 6. In yellow the perimeter of Khanbalik in Beijing.

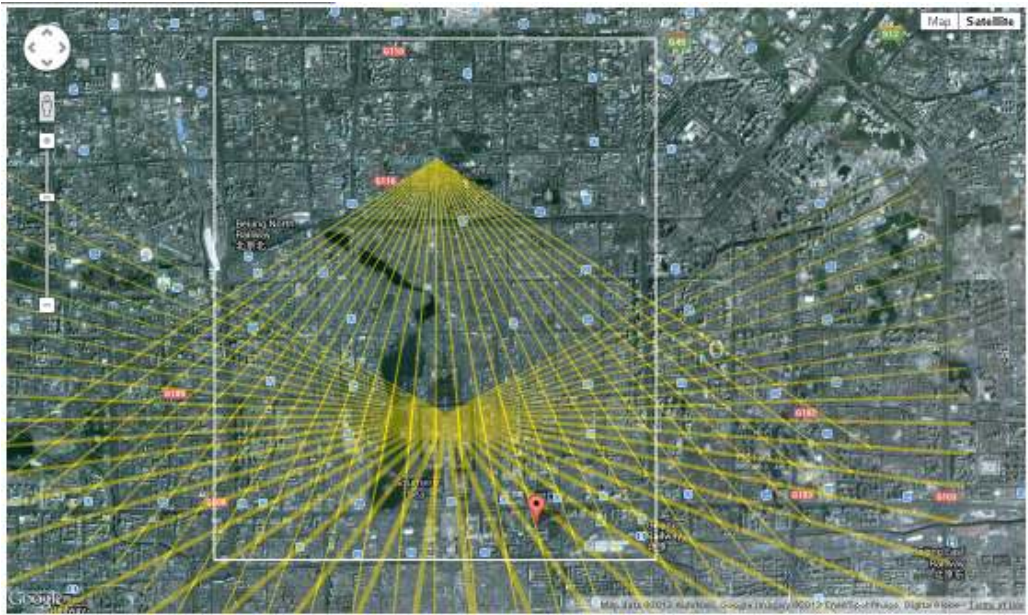

Figure 7: The direction of the sun during the summer and the winter solstices as given by sollumis.com. The polar diagram for the winter solstice has the origin at the crossing of two main roads in the northern part of the city. Using an origin on the Imperial Garden of the Forbidden City, we can have a polar diagram for the summer solstice. The winter and summer sunrise/sunset azimuths cross in two points of the perimeter, which determine a line halving the city.

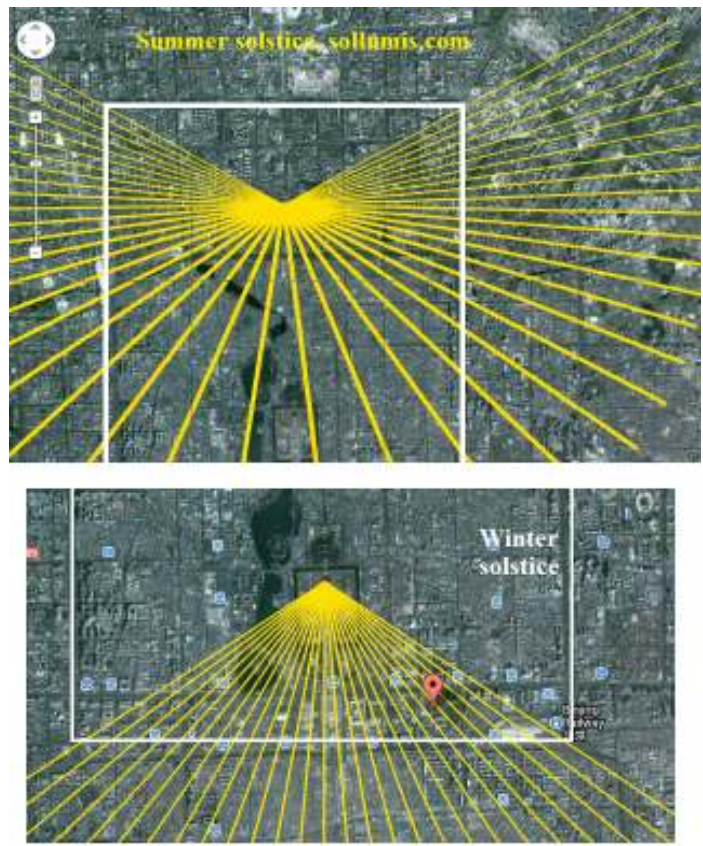

Figure 8: The direction of the sun during the summer/winter solstice. The polar diagrams have the same origins as in Figure 7. The azimuth of sunrise and sunset is passing through the corners of the perimeter.

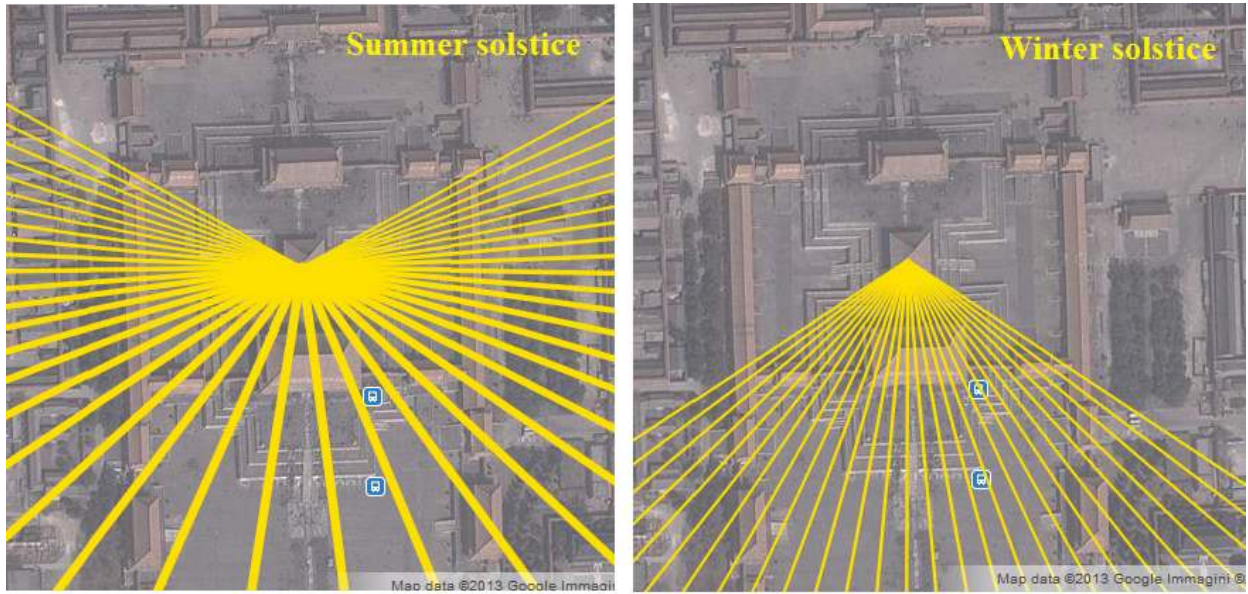

Figure 9: The direction of the sun during the summer/winter solstice as given by sollumis.com. The polar diagrams have origin at the Hall of the Central Harmony in the Forbidden City, which is between the Hall of Preserving Harmony and that of the Supreme Harmony. The azimuths of sunrise and sunset are passing through the corners of the perimeter of the strcture.

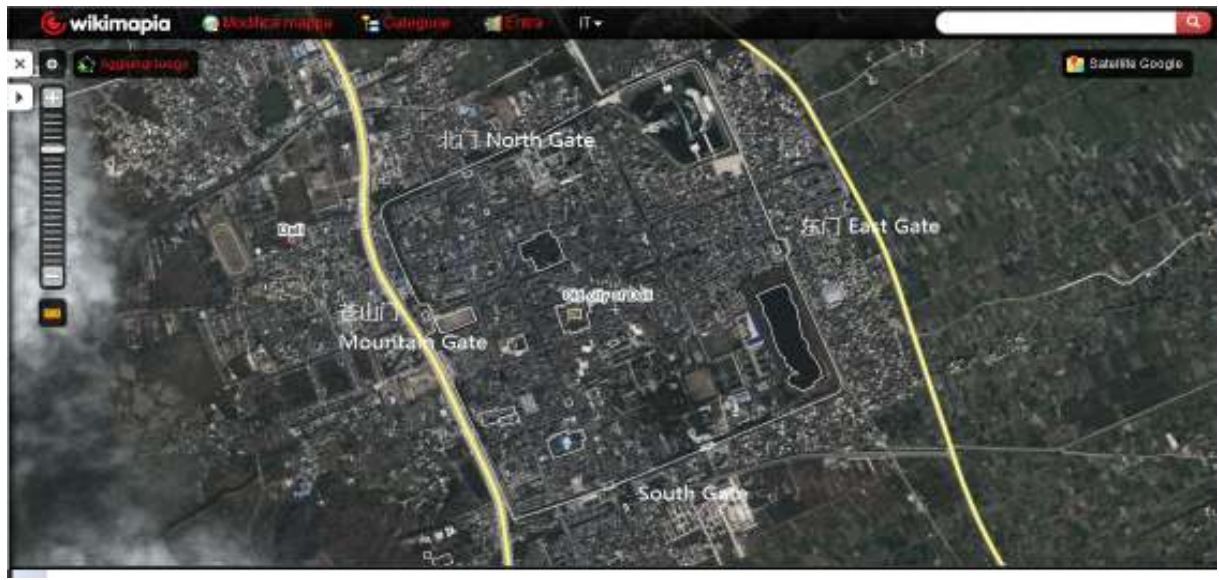

Figure 10: Dali, in the Yunnan province of China, as seen in Wikimapia, with its wall and gates. We have the north and south gates, and the east and the "mountain" gates. Note that the town has not a cardinal orientation.

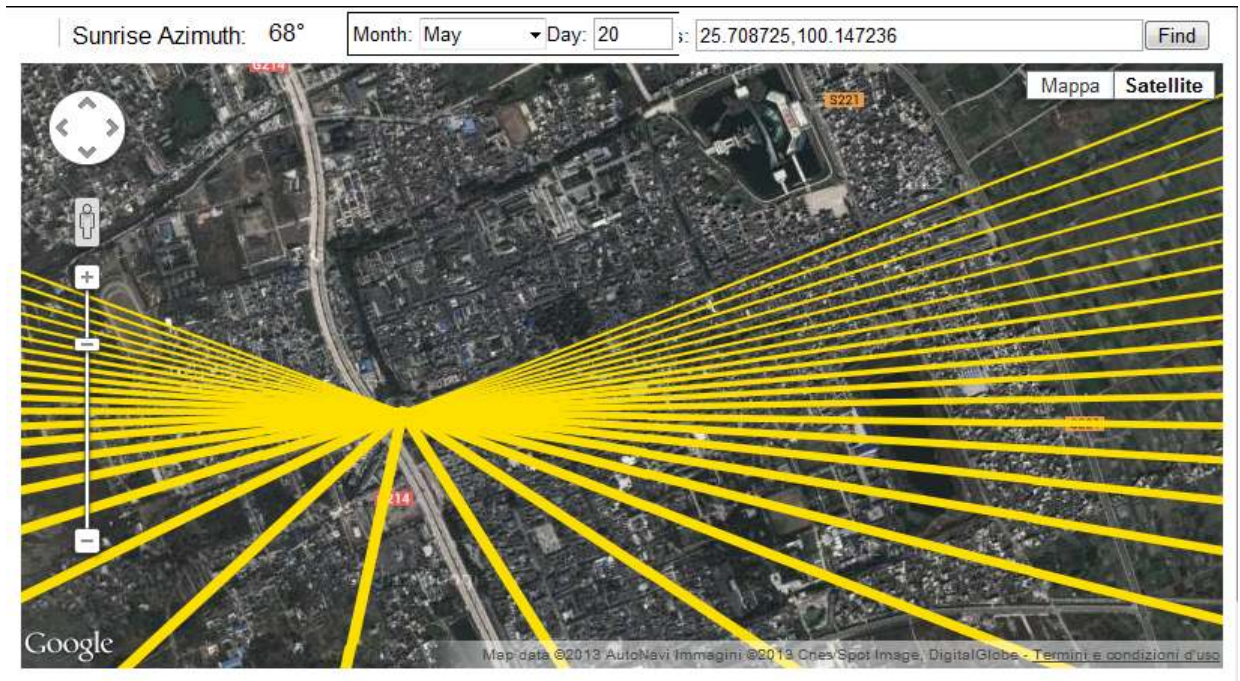

Figure 11: The direction of the sun on 20th of May given by sollumis.com. A solar orientation is then possible.

13